\documentclass[11pt]{article}
\usepackage{epsfig}
\newcommand{\la}[1]{\label{#1}}
\newcommand{\be}{\begin{equation}}
\newcommand{\ee}{\end{equation}}
\newcommand{\ba}{\begin{eqnarray}}
\newcommand{\ea}{\end{eqnarray}}
\newcommand{\bi}{\begin{itemize}}
\newcommand{\ei}{\end{itemize}}

\newcommand{\RR}{{\rm I\kern -.2em  R}}

\newcommand{\helio}{^6{\rm He}}
\newcommand{\neon}{^{18}{\rm Ne}}

\def\lsi{\raise0.3ex\hbox{$<$\kern-0.75em\raise-1.1ex\hbox{$\sim$}}}
\def\gsi{\raise0.3ex\hbox{$>$\kern-0.75em\raise-1.1ex\hbox{$\sim$}}}

\makeatletter \@addtoreset{equation}{section} \makeatother
\renewcommand{\theequation}{\arabic{section}.\arabic{equation}}
\makeatletter
\renewcommand\section{\@startsection {section}{1}{\z@}%
                                   {-5.5ex \@plus -1ex \@minus -.2ex}
                                   {2.3ex \@plus.2ex}%
                                   {\normalfont\large\bfseries}}
\renewcommand\subsection{\@startsection{subsection}{2}{\z@}%
                                     {-3.25ex\@plus -1ex \@minus -.2ex}%
                                     {1.5ex \@plus .2ex}%
                                     {\normalfont\normalsize\bfseries}}
\renewcommand\thesection {\@arabic\c@section}
\renewcommand\thesubsection   {\thesection.\@arabic\c@subsection}
\renewcommand{\@seccntformat}[1]{%
\csname the#1\endcsname.\hspace{1.0em}}
\makeatother

\begin{document}

\begin{titlepage}
\begin{flushright}
December 2003
\end{flushright}

\begin{flushright}
IFIC/03-55      \\
FTUV-03-1205     \\
\end{flushright}

\begin{centering}

\vspace*{0.8cm}

\mbox{\Large\bf 
Neutrino oscillation physics with a higher $\gamma$ $\beta$-beam
}



\vspace*{0.8cm}

J.~Burguet-Castell$^{\rm a,}$\footnote{jordi.burguet-castell@cern.ch}
D.~Casper$^{\rm b,}$\footnote{dcasper@uci.edu}, 
J.J.~G\'omez-Cadenas$^{\rm a,}$\footnote{gomez@mail.cern.ch}, 
P.~Hern\'andez$^{\rm a,}$\footnote{pilar.hernandez@cern.ch}
F.~S\'anchez$^{\rm c,}$\footnote{fsanchez@ifae.es}

\vspace*{0.8cm}

{\em $^{\rm a}$%
Departamento de F\'isica Te\'orica and IFIC, 
Universidad de Val\`encia, 
E-46100 Burjassot, Spain\\}
\vspace{0.3cm}

{\em $^{\rm b}$%
Department of Physics and Astronomy, 
University of California, Irvine
CA 92697-4575, USA\\}

\vspace{0.3cm}

{\em $^{\rm c}$%
IFAE, 
Universidad Autonoma de Barcelona,
E-08193 Bellaterra, Barcelona.\\}
\mbox{\bf Abstract}

\end{centering}

The precision measurement and discovery potential of a neutrino factory based on 
boosted radioactive ions in a storage ring (``$\beta$-beam") is re-examined.
In contrast with past designs, which assume ion $\gamma$ factors 
of $\sim 100$ and baselines of $L=130~\hbox{km}$, we emphasize the
advantages of boosting the ions to higher $\gamma$ and increasing 
the baseline proportionally. In particular, we consider a ``medium-''$\gamma$ scenario ($\gamma \sim 500$, $L\sim 730~\hbox{km}$)  and a ``high-''$\gamma$ scenario
($\gamma \sim 2000$, $L\sim 3000~\hbox{km}$). The increase in statistics, 
which grow linearly with the average beam energy, the ability 
to exploit the energy dependence of the signal and the sizable matter
effects at this longer baseline all increase the discovery potential of such a machine very significantly. 

\vfill

\end{titlepage}

\section{Introduction}

The spectacular results in atmospheric \cite{atmos}, 
solar \cite{solar}, reactor \cite{reactor} and long-baseline \cite{k2k} 
neutrino experiments in recent years can be economically accommodated
in the Standard Model (SM) with neutrino masses and a three-neutrino mixing
matrix \cite{MNS}. In this case, the lepton sector 
of the SM closely resembles that of the quarks and there are a number of 
new physical parameters that can be
measured at low energies: the three neutrino 
masses, $m_i$ ($i=1,2,3$), three mixing angles, $\theta_{ij}$, ($i\neq j = 1,2,3$),  and a CP violating phase, $\delta$. In contrast 
with the quark sector two additional phases could 
be present if neutrinos are Majorana. Of all these new
parameters, present experiments have 
determined just two neutrino mass-square differences and two 
mixing angles: 
($|\Delta m^2_{23}| \simeq 2.5\times 10^{-3}~\hbox{eV}^2$, 
$\theta_{23}\simeq 45^\circ$) which mostly 
drive the atmospheric oscillation and 
($\Delta m^2_{12}\simeq 7\times 10^{-5}~\hbox{eV}^2$,
$\theta_{12}\simeq 35^\circ$) which mostly drive the solar one. 
The third angle, $\theta_{13}$,   as well as the 
CP-violating phases ($\delta$, and possible Majorana phases)
remain undetermined.
Another essential piece of information needed to clarify the 
low-energy structure of the lepton flavor 
sector of the SM is the neutrino mass hierarchy. This is related to 
the sign of the largest mass-square difference ($\Delta m^2_{23}$), which 
determines if the spectrum is hierarchical 
(if the two most degenerate neutrinos are lighter than the third one) 
or degenerate (if they are heavier).

The measurement of  these parameters requires, for the first time,
high-precision neutrino-oscillation experiments. A number
of possible experimental setups to significantly improve the present
sensitivity to $\theta_{13}$, $\delta$ and the sign of $\Delta m^2_{23}$ 
have been discussed in the literature:
neutrino factories (neutrino beams from boosted-muon 
decays) \cite{nufact}, 
superbeams (very intense conventional neutrino beams) 
\cite{JHF,NUMI,Gomez-Cadenas:2001eu}
improved reactor experiments \cite{reactors} 
and more recently $\beta$-beams (neutrinos 
from boosted-ion decays) \cite{zucchelli,mauro}. These are 
quite different in terms of systematics but all 
face a fundamental problem which limits the 
reach of each individual experiment significantly, namely the problem of correlations
and degeneracies between 
parameters \cite{golden}-\cite{silver};
$\theta_{13}$ and $\delta$  
must be measured simultaneously, and 
other oscillation parameters are not known with infinite precision.  

To resolve these degeneracies it is
important to measure as many independent channels as possible and to 
exploit the energy and/or baseline dependence of the 
oscillation signals and matter effects in neutrino propagation. In many 
cases, the best way to do this is by combining different experiments; 
indeed the synergies between some combinations of the setups mentioned above  
have been shown to be very large.

The neutrino factory is thought
to provide ultimate sensitivity to leptonic CP violation,
and thus has been considered for a long time as the last
step on a long-term road map to reveal 
the lepton-flavor sector of the SM. 
In this road map there are important intermediate milestones, such as 
superbeams or improved reactor experiments, that can mitigate the problem of 
degeneracies. In contrast, the present conception of the $\beta$-beam, which 
shares many of the good properties of the neutrino factory, has been shown 
to provide a rather limited sensitivity.

The purpose of this paper is to show that, in fact, a $\beta$-beam
running at a higher $\gamma$ than previously considered (and longer baselines),
in combination with a massive water detector, 
can reach sensitivity to leptonic CP-violation 
and the $\hbox{sign}(\Delta m^2_{23})$ that competes with that in a neutrino 
factory.  On the other hand, the R\&D effort required to increase
the $\gamma$ factor for the $\beta$-beam setup is probably much less than that
required to realize a high-energy, high-luminosity neutrino factory.
Thus, an optimized $\beta$-beam may turn out to be a serious
competitor to the neutrino factory as the ``ultimate machine'' to
search for CP violation.

The paper is organized as follows.
In Section~2 we recall a few facts about the present design for the 
$\beta$-beam and discuss the advantages of increasing the ion $\gamma$ factor 
and the baseline. We also discuss the expected fluxes and 
event rates for three reference setups that will be considered for comparison, 
at low, medium and high $\gamma$. In Section~3, 
the performance of a large water Cerenkov 
apparatus, proposed as the optimal detector for the low and medium
$\gamma$ setups is discussed in detail. Sections~4 and 5 
compare the physics results of the three reference setups. In Section~6 we
present our outlook and conclusions.

%
\section{The $\beta$-beam}

The $\beta$-beam concept was first introduced in~\cite{zucchelli}. 
It involves producing a beam of $\beta$-unstable heavy ions, 
accelerating them
to some reference energy, and allowing 
them to decay in the straight section of
a storage ring, resulting in a very intense neutrino beam.
Two ions have been 
identified as ideal candidates: $^6$He, to produce a pure $\bar{\nu}_e$ beam, 
and $^{18}$Ne, to produce a $\nu_e$ beam. The golden subleading transitions
$\nu_e\rightarrow \nu_\mu$ and ${\bar\nu}_e \rightarrow {\bar\nu}_\mu$ can 
be measured through the appearance of muons in a distant detector.

As in the case of muon-induced neutrino beams, 
the $\beta-$beam
offers the unique features of being {\it pure} (e.g., only one
neutrino species, in contrast to a conventional super-beam
where contamination of other neutrino species is inevitable)
and virtually {\it systematics free}, since the 
spectrum can be calculated exactly
(again, in contrast with a conventional
beam, where knowledge of the spectrum always involves a 
sizable systematic uncertainty). 

One of the most attractive features
of the $\beta$-beam is that it leverages existing 
CERN facilities. The present design,
whose feasability with existing technology has been recently 
demonstrated~\cite{bbcern}, envisions a ``low-''$\gamma$ scenario,
in which ions are produced by a new facility (EURISOL), 
accelerated by the present SPS to $\gamma \leq 150$, and
stored in a storage ring (also a new facility) with straight sections
pointing to the experimental area.
An underground location where a very massive neutrino dectector could be 
located has been identified in the Fr\'ejus tunnel, roughly $130~\hbox{km}$
from CERN. This baseline is ideal for exploring the first
peak of the atmospheric oscillation, the optimal environment to search
for CP-violating effects.   
A new, very large cavern excavated in the Fr\'ejus tunnel would host
a megaton water Cerenkov detector a l\`a UNO. 
The capabilities of such a detector 
are well-matched to this energy range, and the low neutrino energies 
produced by the low-$\gamma$ option (in the range of a few hundred MeV)
require a very large mass to compensate for the tiny cross-sections.

Furthermore, the existing design calls for a conventional
low-energy ``super-beam'' based on the proposed 
SPL proton driver~\cite{splcern}, 
that would deliver a total power-on-target of about
4~MW, resulting in a very intense neutrino beam.
The physics reach of such a super-beam has been studied in detail,
both alone~\cite{Gomez-Cadenas:2001eu}, 
and together with a low-$\gamma$ $\beta-$beam \cite{mauro,blm}.
The results of these studies can be summarized as follows:
\begin{itemize}
\item Neither the SPL super-beam nor the low-$\gamma$ $\beta-$beam, by
themselves, result in fully-satisfactory performance, especially
compared to other proposed facilities such as T2K at JPARC~\cite{JHF}. The
performance is limited, in spite of the large detector mass by 
the small cross-sections, by systematics due to the
SPL super-beam backgrounds (both beam- and detector-related) and by the
intrinsic degeneracies identified in~\cite{burguet1}.
\item A combination of the super-beam and $\beta$-beam 
would explore a large range of the parameter 
space $(\theta_{13},\delta)$. A detailed study of the systematics
involved (in particular the absolute flux normalization of
the SPL super-beam and the uncertainties in background subtraction)
remains to be done, however. On-going studies in the context of the
T2K project suggest that these systematics are not small. 
\item The sign of the atmospheric $\Delta m^2_{23}$ 
cannot be determined because matter effects are negligible.
\end{itemize}

Here, instead, we explore increasing the energy
of the $\beta$-beam, with a corresponding increase in baseline to keep
$\langle E_\nu\rangle/L$ approximately constant. There are three
reasons to expect an improvement of sensitivity in this case:
\begin{itemize}
\item The rates (and thus sensitivity to $\theta_{13}$ if
the backgrounds are kept under control)  increase linearly 
with $\langle E_\nu \rangle$ at fixed $\langle E_\nu\rangle/L$. 
\item At longer baselines, measurement 
of the neutrino mass hierarchy becomes possible, as 
matter effects become more sizable. This is illustrated in Figure~\ref{matter},
which shows the $\nu_e\rightarrow \nu_\mu$ oscillation probability for 
neutrinos and anti-neutrinos, as a function 
of the baseline, for neutrino energy 
constrained to the first atmospheric peak, i.e. 
$E/L = |\Delta m^2_{23}|/2\pi$. The difference between the neutrino and 
anti-neutrino oscillation probabilities induced by matter effects becomes 
comparable to that due to CP-violation for $L={\cal O}(1000)km$. 
\item Increased neutrino energy enhances the energy dependence of 
oscillation signals. This is extremely 
useful in resolving the correlations and degeneracies 
in parameter space~\cite{golden,burguet1}. 
Energy dependence
is particularly helpful for energies close to the peak of the atmospheric 
oscillation \cite{burguet2}, 
which is precisely the regime under consideration.  
Figure~\ref{degen} 
compares the vacuum probabilities for 
$\nu_e\rightarrow\nu_\mu$  and ${\bar\nu}_e\rightarrow{\bar\nu}_\mu$ with  
$\theta_{13}=6^\circ$ and $\delta=40^\circ$ to those of
the intrinsic-degenerate solution at $\langle E_{\nu({\bar\nu})} \rangle$. 
The neutrino and anti-neutrino probabilities cross at 
the $\langle E_{\nu({\bar\nu})} \rangle$, but are quite 
different at other energies. Thus spectral  
information can definitely help in disentangling them. 
\end{itemize}

\begin{figure}[t]
\begin{center}

\epsfig{file=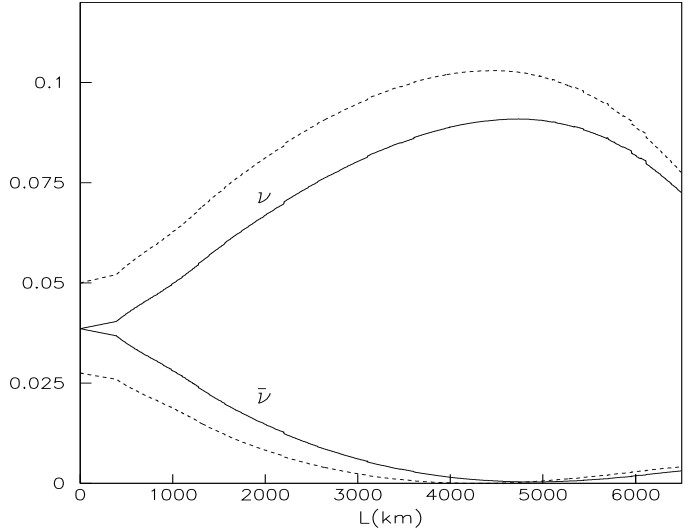,width=10cm,height=8cm} 

\caption[a]{$P(\nu_e\rightarrow\nu_\mu)$ and $P(\bar\nu_e\rightarrow\bar\nu_\mu)$ 
as a function of the baseline $L$ in km, at a neutrino energy 
$E/L = |\Delta m^2_{23}|/2\pi$ and 
for $\theta_{13} =8^\circ$ and $\delta=0$ (solid) and $90^\circ$ (dashed). }
\la{matter}

\end{center}
\end{figure}

\begin{figure}[t]
\begin{center}

\epsfig{file=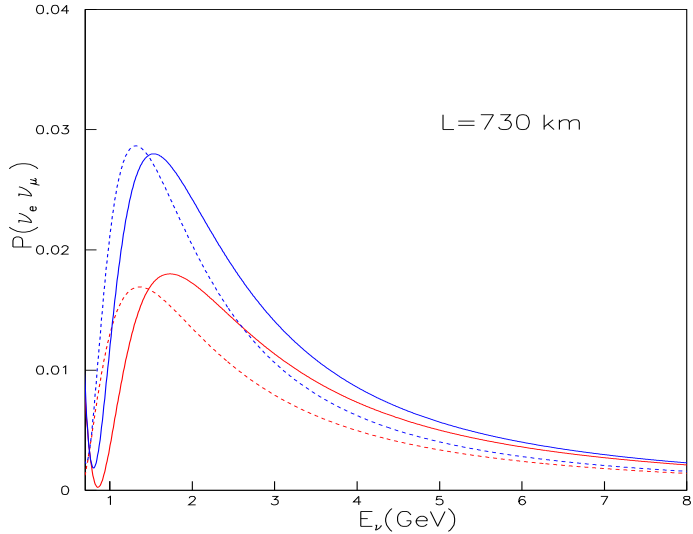,width=10cm,height=8cm} 

\caption[a]{$P(\nu_e\rightarrow\nu_\mu)$ 
and $P(\bar\nu_e\rightarrow\bar\nu_\mu)$ as a 
function of the energy in GeV, at $L=730$~km, for $\theta_{13}=6^\circ$ 
and $\delta=40^\circ$ (solid) and for the degenerate solution 
at $\theta'_{13}(\langle E_\nu \rangle), \delta'(\langle E_\nu \rangle)$ 
with $\langle E_\nu \rangle =1.5~\hbox{GeV}$ (dashed). } 
\la{degen}

\end{center}
\end{figure}

From a technical point of view, designs aiming at higher $\gamma$ factors 
are conceivable by direct extrapolation of existing technology, and would
not require a long R\&D program.  A ``medium-''$\gamma$
scenario ($\gamma \leq 600$) could be realized at CERN by accelerating
ions in a refurbished SPS with super-conducting magnets, or in LHC  
(up to $\gamma =$2488 for $\helio$ and $\gamma = 4158$ for $\neon$)~\cite{matsmo}. 
Another candidate would be Fermilab, where a combination of the existing
Main-Injector and Tevatron could accelerate ions to 
$\gamma$ factors of a few hundred. 


The possibility of an associated
super-beam\footnote{Note that it is also possible to increase the
energy of the super-beam in the present design, by incresing the
energy of the SPL.} will not be considered here, because the systematics
are very different. To gain a clear view of the
relative merits of each $\beta-$beam 
scenario, it is best to first compare their
stand-alone performance.  Three scenarios for the $\beta-$beam {\it
alone} are considered, and a future paper will address comparisons
and combinations with other facilities.  We define the following three setups:
\begin{itemize}
\item Setup I, low energy: $\gamma=60$ for $\helio$ and 
$\gamma=100$ for $\neon$, with $L=130~\hbox{km}$ (CERN--Fr\'ejus) 
as in~\cite{mauro,blm}. 
\footnote{Different $\gamma$ 
for $\helio$ and $\neon$ are required to allow simultaneous 
acceleration and storage of both
ions in the same ring,
reducing the necessary running time by a
factor of two. The different ion charge:mass ratios 
imply a $1.67$ ratio of $\gamma$ factors~\cite{matsmo}.}
\item Setup II, medium energy: $\gamma=350$ for $\helio$ and
$\gamma=580$ for $\neon$, with $L=732~\hbox{km}$ (e.g. CERN--Gran Sasso with
a refurbished SPS or with the LHC, FNAL--Soudan with Tevatron).
\item Setup III, high energy: 
$\gamma=1500$ for $\helio$ and $\gamma=2500$ for $\neon$,
with $L=3000~\hbox{km}$ (e.g. CERN--Canary islands with the LHC). 
\end{itemize}

In all cases we assume the same number of ions as the
existing design, that is, $2.9\times 10^{18}~\helio$ and 
$1.1\times 10^{18}~\neon$ decays per year 
\footnote{Although the originally-projected intensity 
of $\neon$ was a factor three less, it was recently proposed~\cite{matsmo} 
that three bunches could be accommodated.}. This seems reasonable, as one
does not expect loses with a refurbished SPS (to extrapolate,
for instance, setup-I to setup-II at CERN). For the LHC one
could compensate for injection loses due to the different optics with
a different acceleration scheme with more or longer 
bunches (thus more ions)~\cite{matspc}. 

%
\subsection{Neutrino fluxes}

Neglecting small Coulomb corrections, the electron energy spectrum
produced by an ``allowed" nuclear $\beta-$decay
at rest is described by:
\be
{dN^{\rm rest} \over d p_e }\sim p_e^2 (E_e-E_0)^2, 
\ee
where $E_0$ is the electron end-point energy and $E_e$ and $p_e$ are the 
electron energy and momentum. For $^6$He, $E_0=3.5~\hbox{MeV}+m_e$, while for 
$^{18}$Ne, $E_0=3.4~\hbox{MeV}+m_e$.

We are interested instead in the neutrino spectrum resulting from ion decays 
after they are boosted by some fixed $\gamma$. In the ion rest frame the spectrum of 
the neutrinos is 
\be
{dN^{\rm rest} \over d \cos\theta d E_\nu} \sim E_\nu^2  (E_0-E_\nu) \sqrt{(E_\nu-E_0)^2-m_e^2}.
\ee

After performing the boost and normalizing the total number of ion decays to be
$N_\beta$ per year, the neutrino flux per solid angle in a detector located at a distance $L$ 
aligned with the straight sections of the storage ring  is:
\ba
\left.{d\Phi^{\rm lab}\over dS dy}\right|_{\theta\simeq 0} 
\simeq {N_\beta \over \pi L^2} {\gamma^2 \over g(y_e)} y^2 (1-y) \sqrt{(1-y)^2 - y_e^2}, 
\ea
where $0 \leq y={E_\nu \over 2 \gamma E_0} \leq 1-y_e$, $y_e=m_e/E_0$ and 

\be
g(y_e)\equiv {1\over 60} \left\{ \sqrt{1-y_e^2} (2-9 y_e^2 - 8 y_e^4) + 15 y_e^4 Log\left[{y_e \over 1-\sqrt{1-y_e^2}}\right]\right\}
\ee
Note the similarity of this expression and the electron neutrino 
fluxes at a neutrino factory~\cite{nufact}.  Another similarity is that 
the fluxes are known very accurately and the $\nu_\mu(\bar{\nu}_\mu)$ appearance signal
has no background from contamination of the beam.  The latter is true for the neutrino 
factory only to the extent that 
the charge of final-state leptons can be determined, 
which requires, therefore, a magnetized device (thus, in 
particular it prevents the use of massive water detectors). 

Figure~\ref{fig:fluxes} shows these fluxes at the three 
reference setups as a function of the neutrino energy.
Although integrated fluxes for all the three 
setups are nearly identical, 
about ${\cal O}(10^{11}) \bar{\nu}_e/\nu_e~ m^{-2}~\hbox{year}^{-1}$, 
setups~II and III have the advantage that
the appearance signal's 
energy dependence should be more significant, while at low energy 
the neutrino energy resolution is degraded by the Fermi motion. 

\begin{figure}[t]
\begin{center}

\epsfig{file=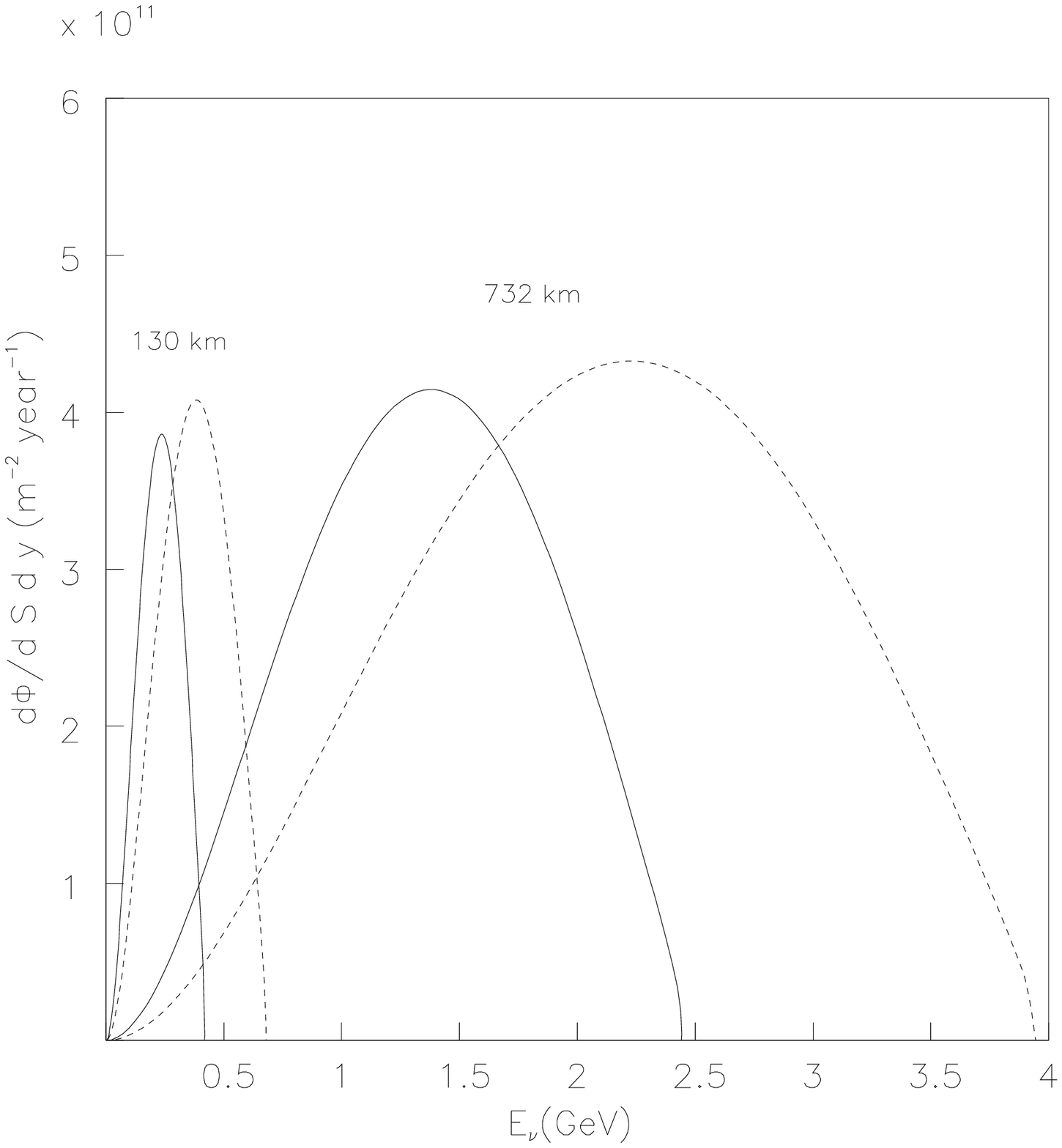,width=7.5cm,height=8cm} \epsfig{file=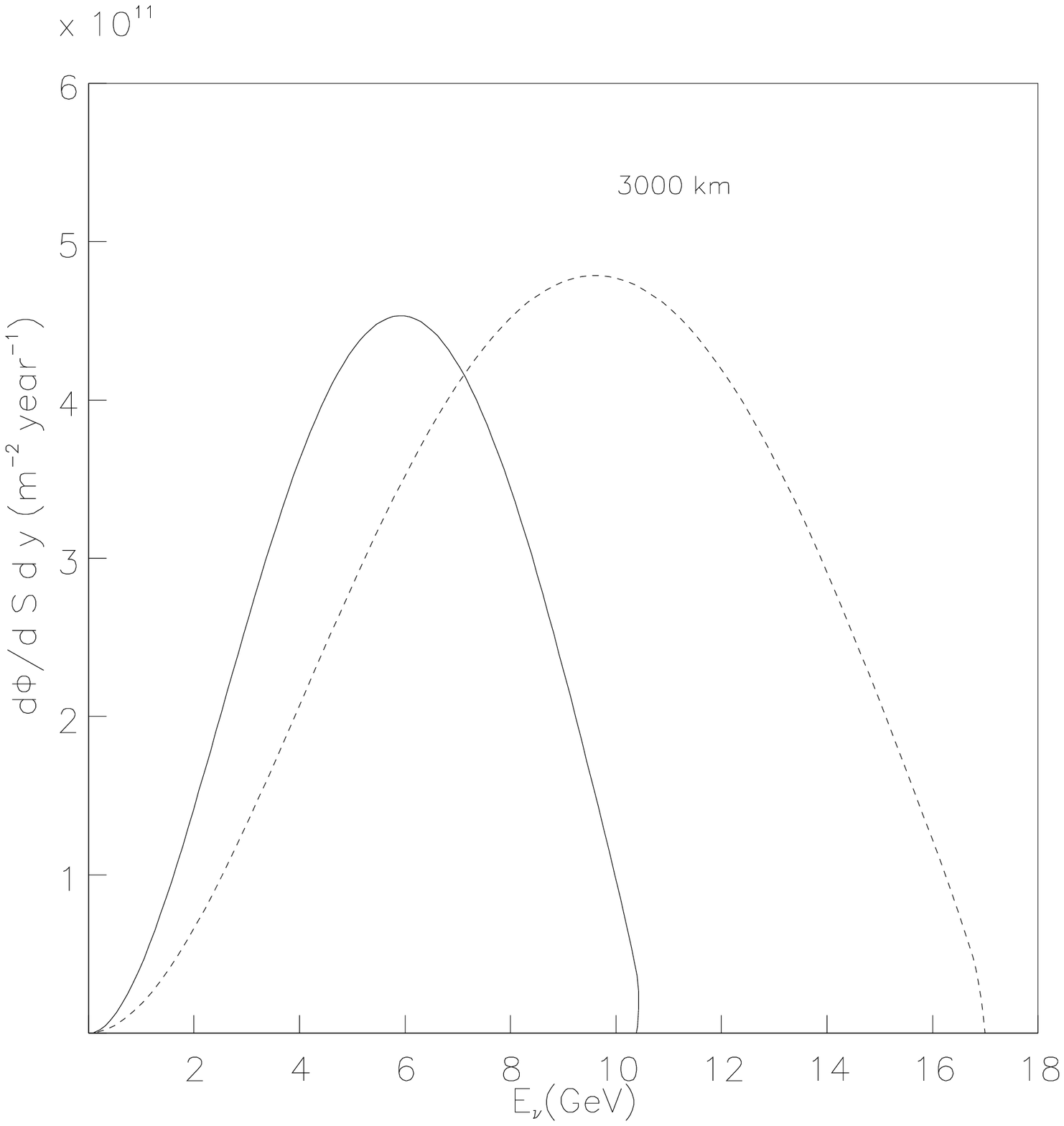,width=7.5cm,height=8cm} 

\caption[a]{Fluxes for the three setups as function of the neutrino energy
for $\bar{\nu}_e$ (solid) and $\nu_e$ (dashed).}
\la{fig:fluxes}

\end{center}
\end{figure}

It is instructive to compare with the neutrino 
factory flux of $10^{12}~ \bar{\nu}_e/\nu_e~ m^{-2}~\hbox{year}^{-1}$ 
at the optimum baseline of $3000~\hbox{km}$.  This is a 
factor 10 higher than setup~II, but $\langle E_\nu\rangle/L$ 
for setups~II and III is matched to the atmospheric 
splitting, while at the neutrino factory it is not. The oscillation probabilities are
thus smaller in the latter.

\subsection{Cross-sections}

The (anti-)neutrino cross-sections relevant in the three setups are 
quite different. While in the low energy option quasi-elastic events
are dominant and the cross-section grows rapidly with energy, 
in the highest-energy option samples are mostly deep-inelastic scattering and the 
growth is linear in the neutrino energy. For the medium-energy
option, there is a sizable contribution from both types of events, as well as resonant channels. 
Figure~\ref{fig:xsec} shows the cross-sections 
per nucleon and per neutrino energy used
in this analysis. 

\begin{figure}[t]
\begin{center}

\epsfig{file=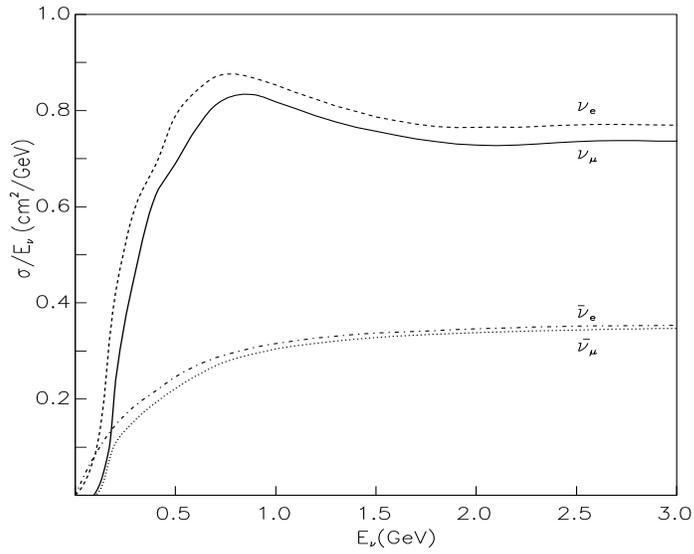,width=10cm,height=8cm} 

\caption[a]{Cross section per nucleon for an isoscalar target, divided by neutrino energy in GeV.}
\la{fig:xsec}

\end{center}
\end{figure}

Asymptotically the number
 of events grows linearly with $\gamma$. Table ~1
shows the number of charged-current events expected in one year, per kiloton.  The number
of events is somewhat sensitive to the isotopic composition of the target, as free protons in water
contribute significantly to the anti-neutrino event rates.

\begin{table}
\begin{center}
\begin{tabular}{|c|c|c|c|c|}
\hline
$\gamma$ & $L(km)$ & ${\bar\nu}_e$ CC & $\nu_e$ CC & $\langle E_\nu \rangle (GeV)$\\
\hline
60/100 & 130 & 4.7 & 32.8 & 0.23/0.37 \\
\hline
350/580 & 730 & 57.5 & 224.7 & 1.35/2.18\\
\hline
1500/2500 & 3000 & 282.7 & 993.1 & 5.80/9.39\\
\hline
\end{tabular}
\label{table:CCevents}
\caption{Number of charged-current events without oscillations per \hbox{kton-year} 
for the three reference setups. Also shown is the average neutrino energy.}
\end{center}
\end{table}

Interestingly, the detector mass can be reduced 
linearly with $\gamma$ without changing the number of events. This offers 
the possibility of moving from the large water Cherenkov detector required
in the lowest-energy option to a less massive but more granular detector 
in the higher-energy ones.

%
\section{Detectors: efficiencies and backgrounds}
\label{detectors}

The signature for the golden subleading transitions
$\nu_e\rightarrow \nu_\mu$ and ${\bar\nu}_e \rightarrow {\bar\nu}_\mu$ in
a~$\beta$-beam is the appearance of prompt muons which must be
separated from the main background of prompt electrons due to 
the bulk $\nu_e/{\bar\nu}_e$ charged interactions. Since there is
only one neutrino species in the beam, no charge 
identification is required. Furthermore,
to compensate the small cross sections, specially for setup-I, very massive
detectors are needed. In addition, 
good energy resolution is required in order to resolve parameter degeneracies.

As demonstrated by Super-Kamiokande \cite{atmos}, massive water 
detectors are capable 
of offering simultaneoulsy excellent particle identification and good
energy resolution, particularly  in the range 
of few hundred MeV to about
1 GeV, where most of the interactions are 
quasi elastic, yielding simple event topologies
(a typical QE interaccion is characterized by a single
ring from the final muon, the scattered proton 
being below Cerenkov threshold
thus invisible). As energy increases, deep inelastic 
processes start to dominate
the cross section and the event topology becomes 
more complicated. The turn-over 
region is about 1.5 GeV. The neutrino spectra in Setup-II extend all the way up
to 4 GeV. Nevertheless, as it will be shown below, water is still 
the best option in this range. 

For neutrinos energies up to 10 GeV, as considered in 
setup-III, deep inelastic CC and NC events are dominant and water is no
longer suitable. Massive tracking calorimeters
are the best option in this range \cite{cervera,golden}. 

\subsection{Signal selection and background supression in water}

\label{uno}

We have considered a Megaton-class water detector, as proposed
by the UNO collaboration \cite{uno}
with a fiducial mass of 400 kiloton, for both setups I and II. The
response of the detector was studied using the NUANCE \cite{nuance} neutrino
physics generator and detector simulation and realistic reconstruction
algorithms as described in \cite{Gomez-Cadenas:2001eu}.

Particle identification in water exploits the difference in the Cerenkov
patterns  produced by showering (``e-like") and non-showering
(``$\mu$-like") particles. Besides, for the energies of interest
the difference in Cerenkov opening angle between an
electron and a muon can also be exploited. Furthermore, muons
which stop and decay (100\% of $\mu^+$ and 78\% of $\mu^-$)
produce a detectable delayed electron signature that can be used
as an additional handle for background rejection.

\begin{figure}[htb]
\centerline{\epsfig{file=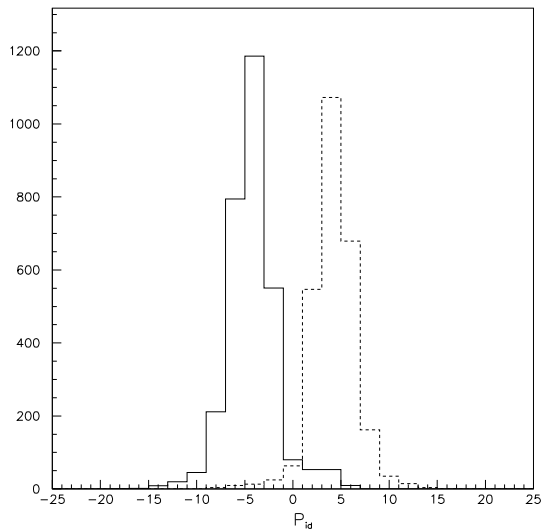,width=8.0cm}}
\caption{Rejection of  $\nu_{e}^{CC}$ background
in a water Cerenkov detector. The particle ID estimator $P_{\rm id}$ (in
arbitrary units) is shown for the electron-like background (left, solid
line) and muon-like signal (right, dashed). }
\label{fig:pid}
\end{figure}

For this study, we have used a particle
identification criteria similar to the one
used by the Super-Kamiokande collaboration, which is based 
on a maximum likelihood
fit of both $\mu$-like and e-like hypotheses. 
Figure~\ref{fig:pid} shows the particle identification estimator $P_{id}$, 
for electron-like events (solid line) and for muon-like events (dashed line).
The normalization is arbitrary. A
cut at $P_{id}>0$ (PID cut) separates optimally the e-like and $\mu$-like populations.
Since most $\nu_{\mu}$ 
events are followed
by a muon-decay signature, the background is further reduced by
accepting only events with a delayed coincidence. 

To summarize, the event selection for both setup-I and setup-II requires:
\begin{itemize}
\item The event must be fully contained in the fiducial volume. 
This is necessary to guarantee a
good measurement of the energy as well as to exploit the muon-decay signature.
\item A single ring in the event.
\item The PID estimator must be muon-like $P>0$.
\item Event must show a delayed coincidence (muon decay signature).
\end{itemize}

\subsection{Signal and backgrounds in setup-I}

The PID cut eliminates almost completely the electron background,
leaving a residual background due to $\nu_{e}^{NC}$ and diffractive 
events in which a single pion is confused with a muon. 
The low energy of setup-I (particularly in the case of the 
antineutrinos) results in negligible backgrounds for $\helio$. In the
case of $\neon$, diffractive events result in an
 integrated background fraction below 
$10^{-2}$. The efficiency is rather large but drops dramatically below 
300~MeV. Background ratio
and efficiency as a function of the energy in setup-I are shown
in Figure~\ref{fig:eff_low}.

\begin{figure}[t]

\epsfig{file=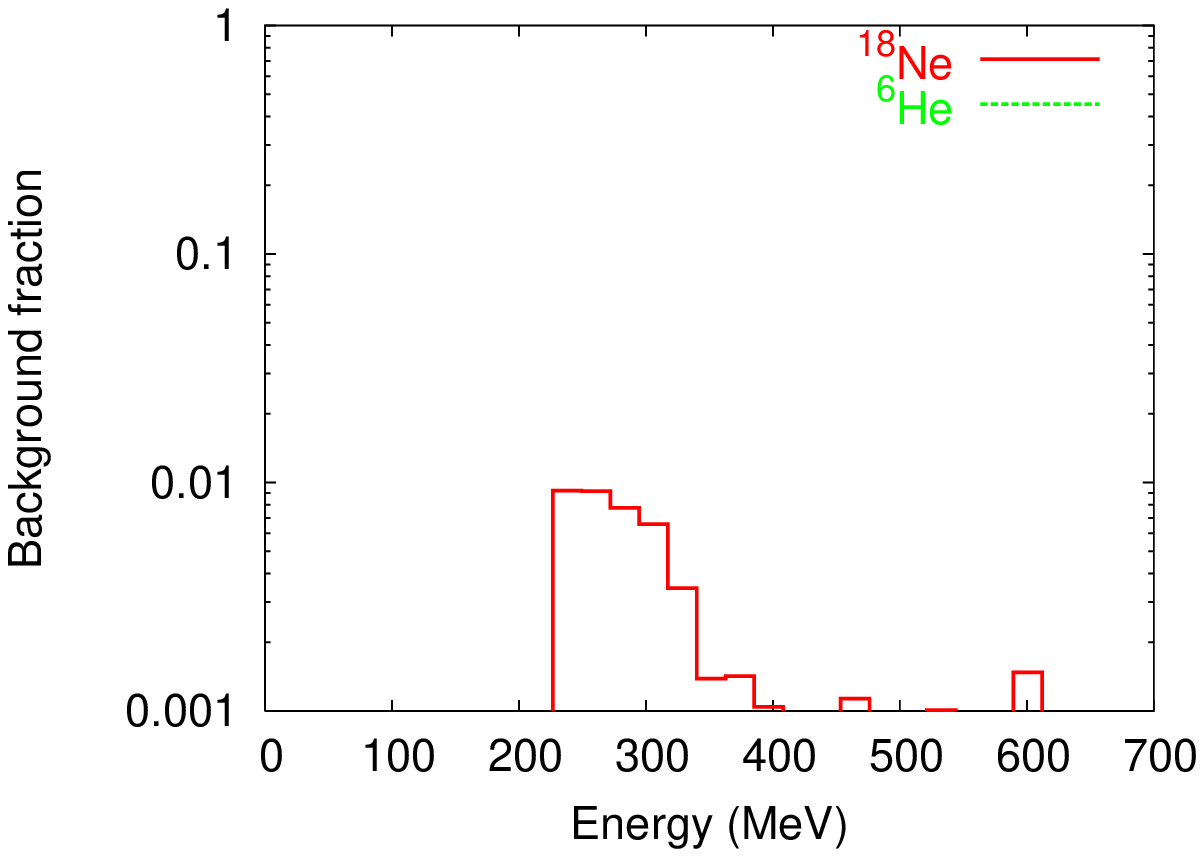,width=7.5cm,height=7cm} 
\epsfig{file=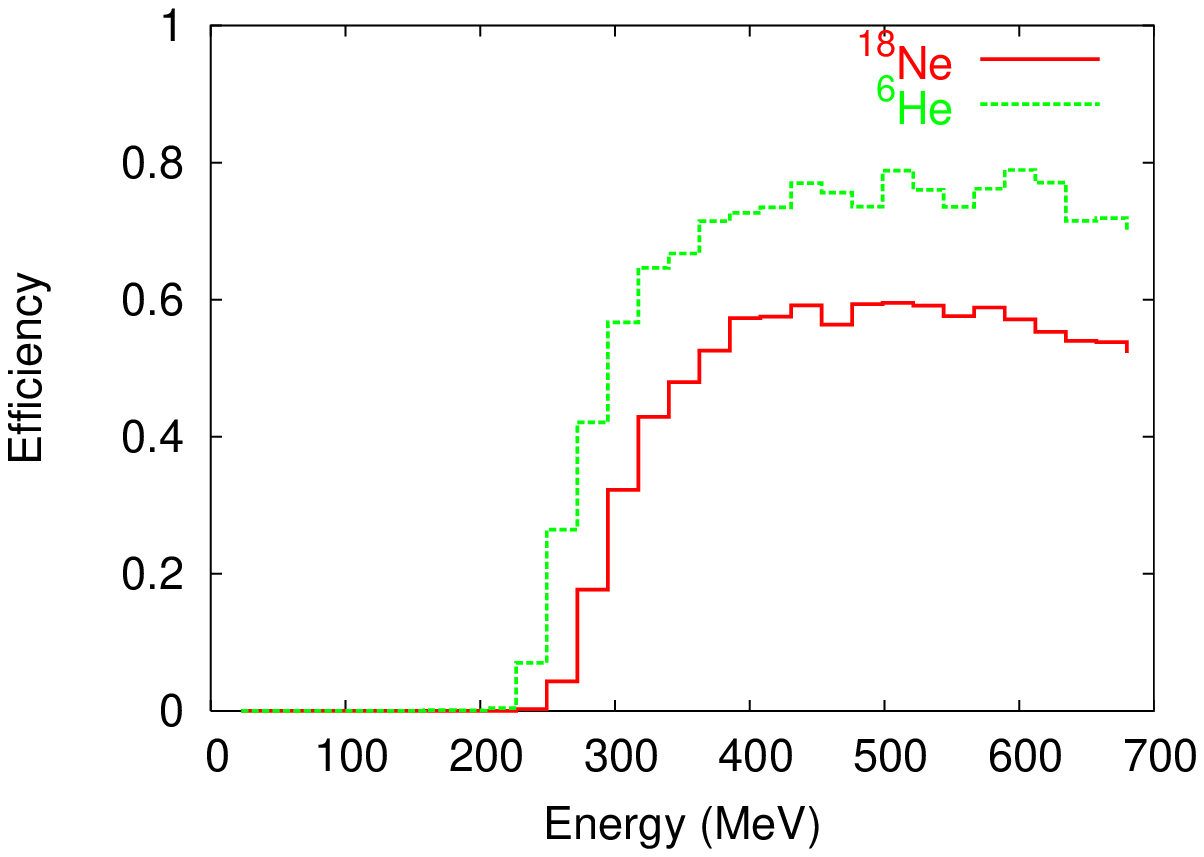,width=7.5cm,height=7cm} 

\caption{ Background fraction (left) and efficiency (right) as a function of the true 
neutrino energies for $^6$He and $^{18}$Ne in water in Setup I.}
\la{fig:eff_low}

\end{figure}

\begin{figure}[htb]
\centerline{\epsfig{file=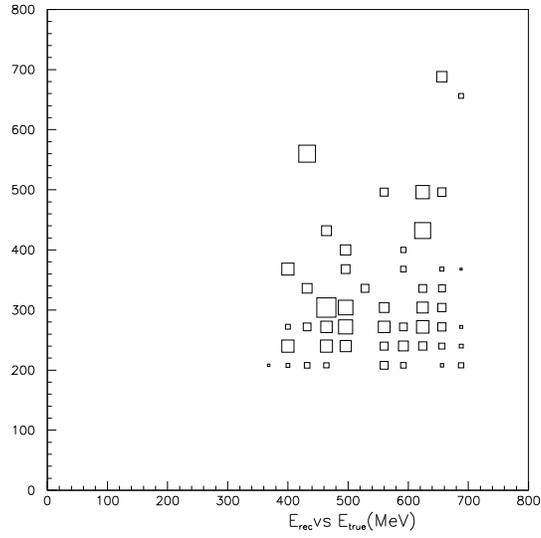,width=8.0cm}}
\caption{Reconstructed versus true
neutrino energy for $^{18}$Ne. The lack of correlation shows
clearly that the event energy information is completely
washed out by Fermi motion.} \label{fig:fermi}
\end{figure}

A major drawback of setup-I is that no energy binning is possible,
since the neutrino energy is of the order of the Fermi motion.
This is illustrated in Figure \ref{fig:fermi} where the reconstructed
neutrino energy is plotted against the true energy. As it can be
seen the events are almost uncorrelated. Therefore, spectral information
cannot be used and one has to make do with the integrated signal, which,
alas, cannot resolve the intrinsic degeneracies.

\subsection{Signal and backgrounds in setup-II}

\begin{figure}[t]
\epsfig{file=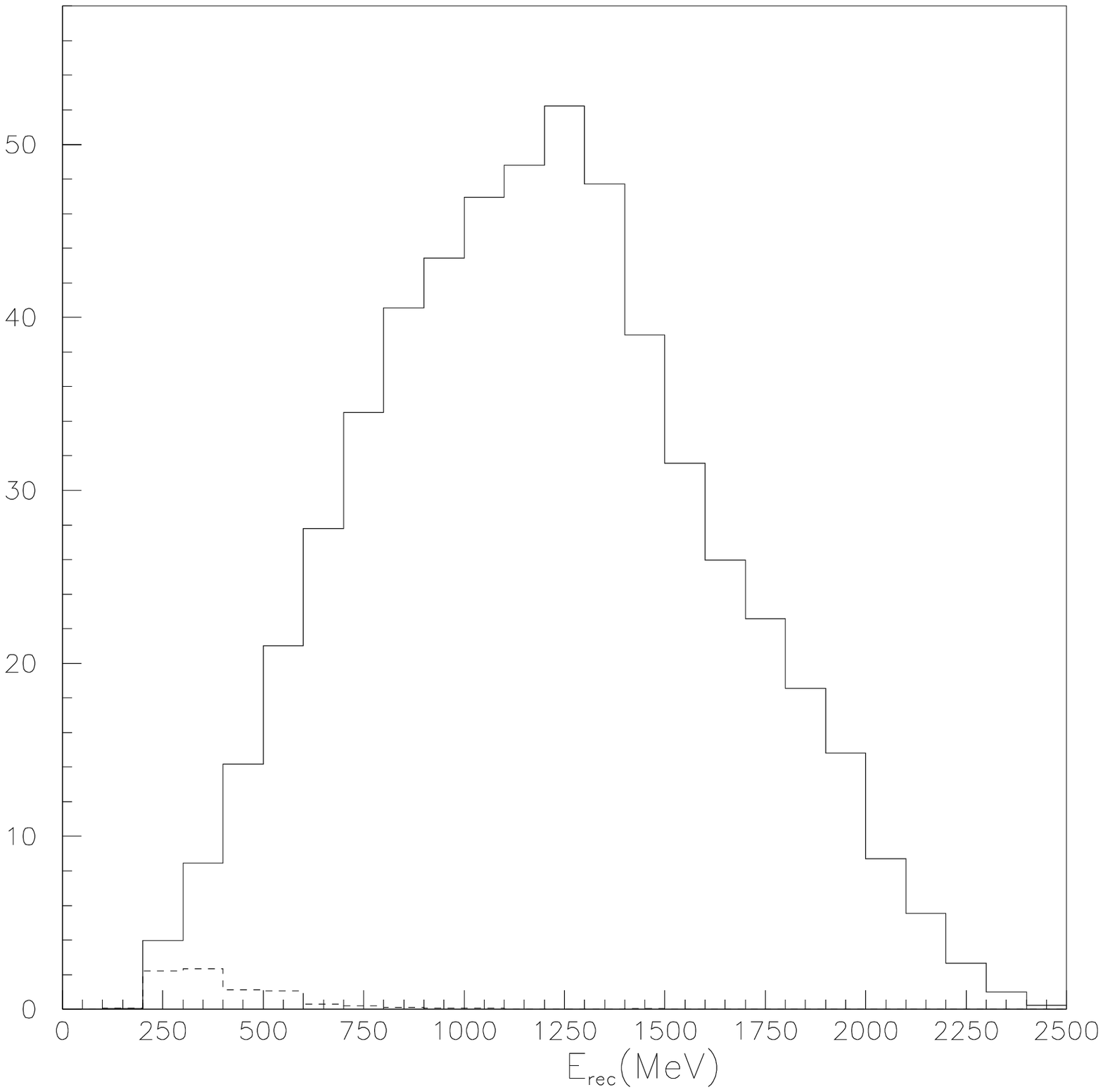,width=7.5cm,height=7.5cm} 
\epsfig{file=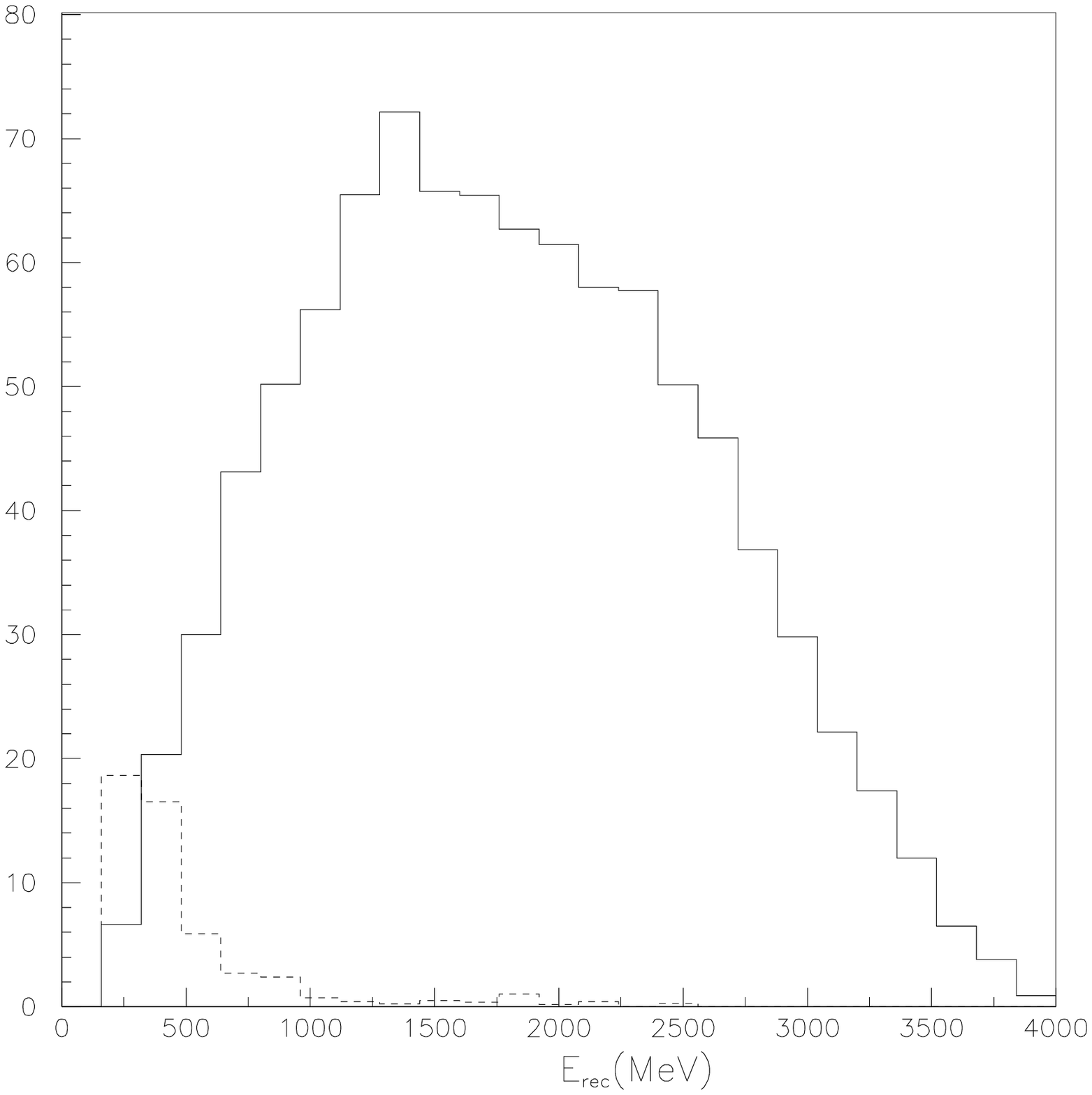,width=7.5cm,height=7.5cm} 

\caption{Reconstructed energy for signal and background in Setup-II 
(the absolute normalization is arbitrary) for $^{6}$He (left) and $^{18}$Ne (right).}
\la{fig:erec}

\end{figure}
Figure \ref{fig:erec} shows the reconstructed energy spectrum for
both signal and background in setup-II. Notice
that, as for Setup-I, the backgrounds are smaller for
$^{6}$He than for $^{18}$Ne, and that both neutrino and antineutrino
backgrounds tend to cluster at low energies. A cut demanding that the
reconstructed energy be larger than 500 MeV supresses most of the
residual backgrounds at a modest cost for the efficiency.

\begin{figure}[t]
\epsfig{file=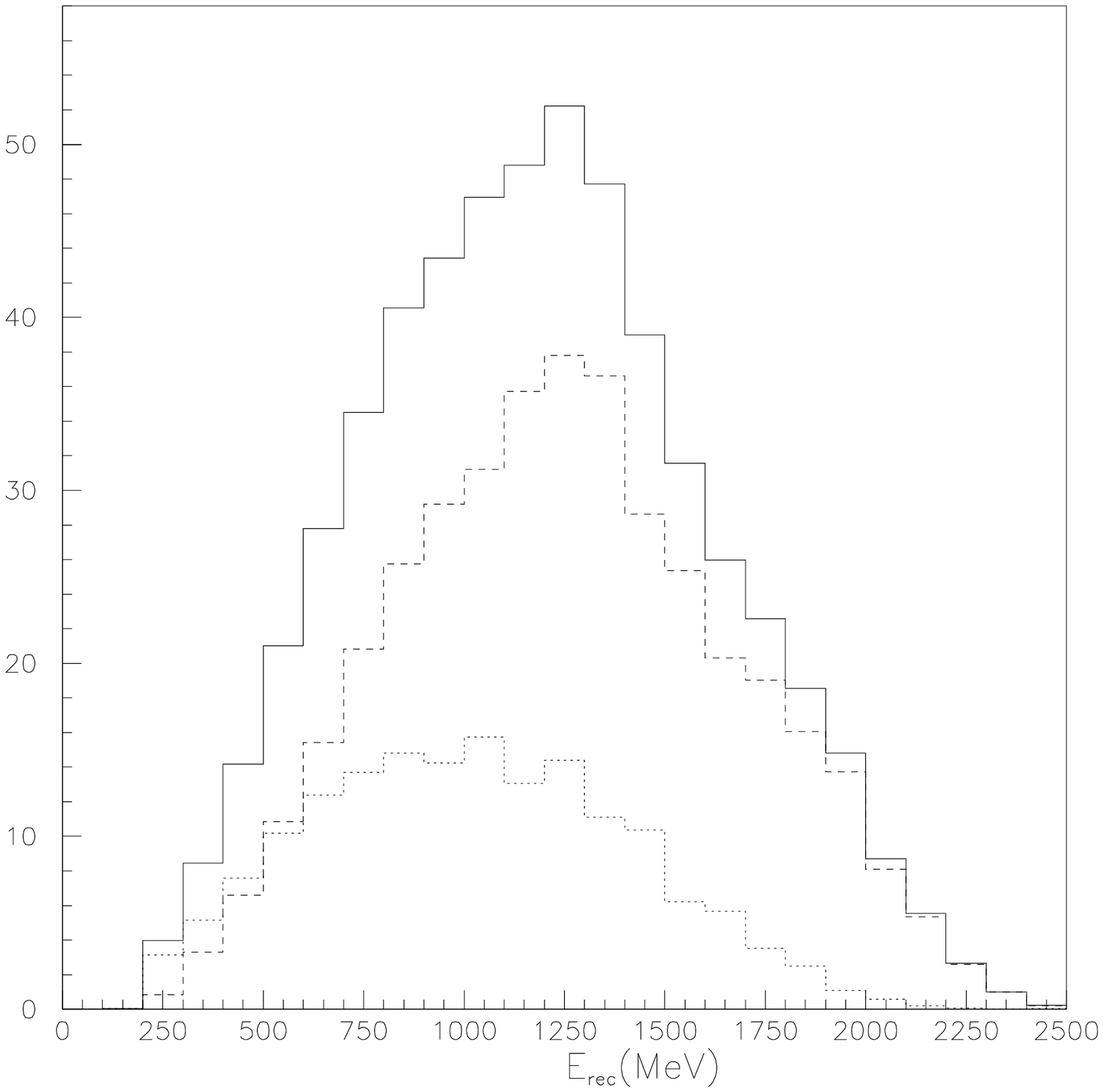,width=7cm,height=7cm} 
\epsfig{file=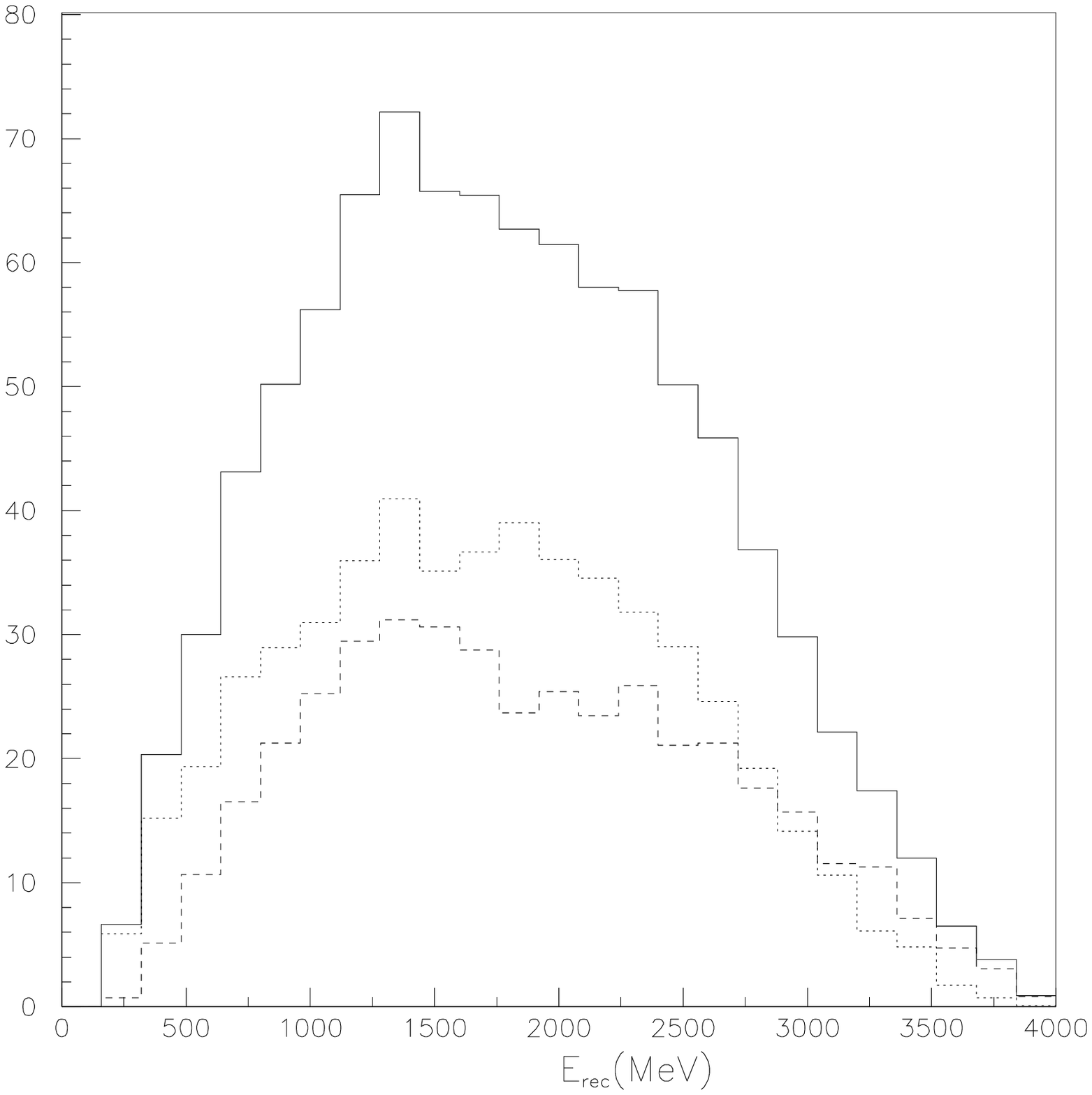,width=7cm,height=7cm} 

\caption[a]{
Reconstructed energy (solid line), the QE component (dotted line)
and the non-QE component (dashed line) for signal-like events. 
(arbitrary normalization) for $^{6}$He (left) and $^{18}$Ne (right).}
\la{fig:erec2}

\end{figure}

Figure \ref{fig:erec2} shows
the reconstructed energy (solid line), the QE component (dotted line)
and the non-QE component (dashed line) for events passing the
selection criteria.
Notice that the non-QE contamination is high, specially for neutrino
events. This spoils sizeably the resolution, since the
neutrino energy is reconstructed under the hypothesis that the interaction
was QE. The effect is ilustrated in Figure~\ref{fig:eres}, which shows
the reconstructed versus true energy for antineutrinos in
Setup-II for QE events only (left) and 
all events (QE and non QE) 
that pass the selection criteria (right). 
Notice the excelent correlation between reconstructed and
true energy in the case of QE events, which is, however,
spoiled by the non-QE events. 
We take into account this effect by computing a matrix
that describes the migrations between the true and
the reconstructed neutrino energy. Migration matrices have also
been computed for the backgrounds. 
Figures ~\ref{fig:effbgskhe6} and ~\ref{fig:effbgskne18}  show those 
matrices (in the form of lego-plots) for $\helio$ and $\neon$ 
respectively. The integrated efficiencies are quite high 
($\sim 30-50\%$) for background fractions below $3 \times 10^{-3}$.

\begin{figure}[t]
\epsfig{file=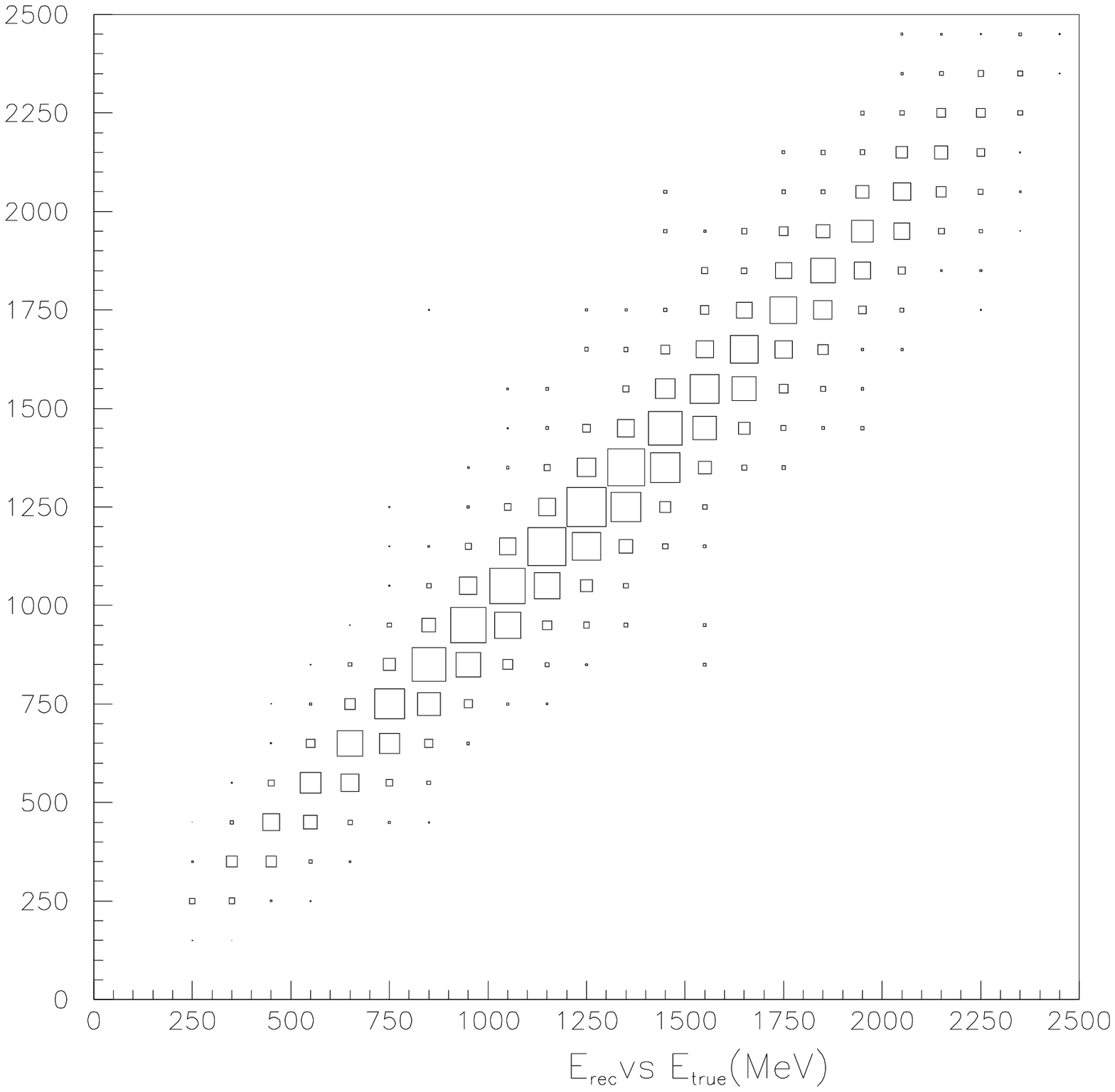,width=7.5cm,height=7.5cm} 
\epsfig{file=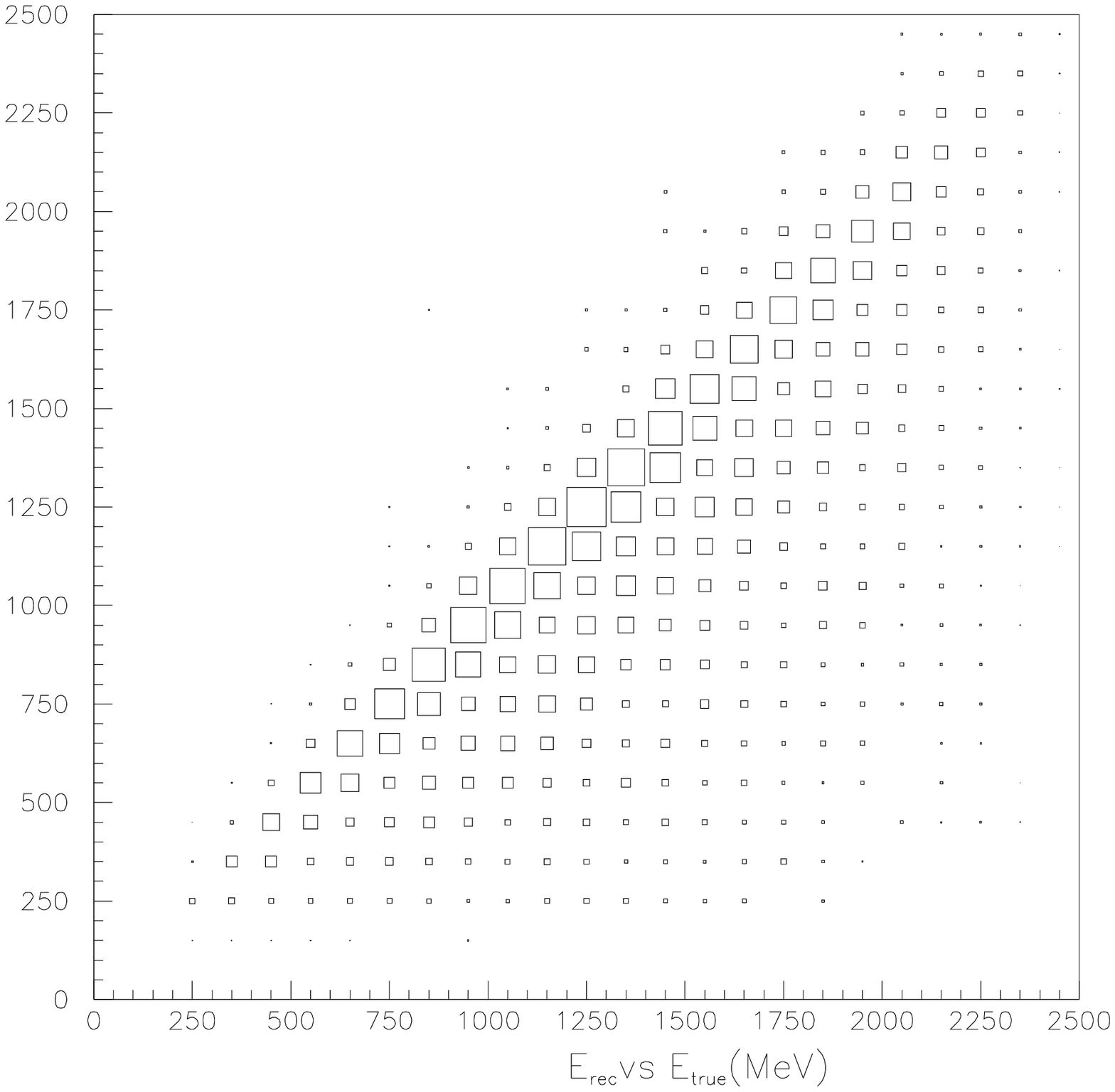,width=7.5cm,height=7.5cm}

\caption{Reconstructed versus true energy for antineutrinos in
Setup-II for  QE events on the left and for
signal-like events (QE and non-QE events) on the right.}
\la{fig:eres}

\end{figure}

\begin{figure}[t]

\epsfig{file=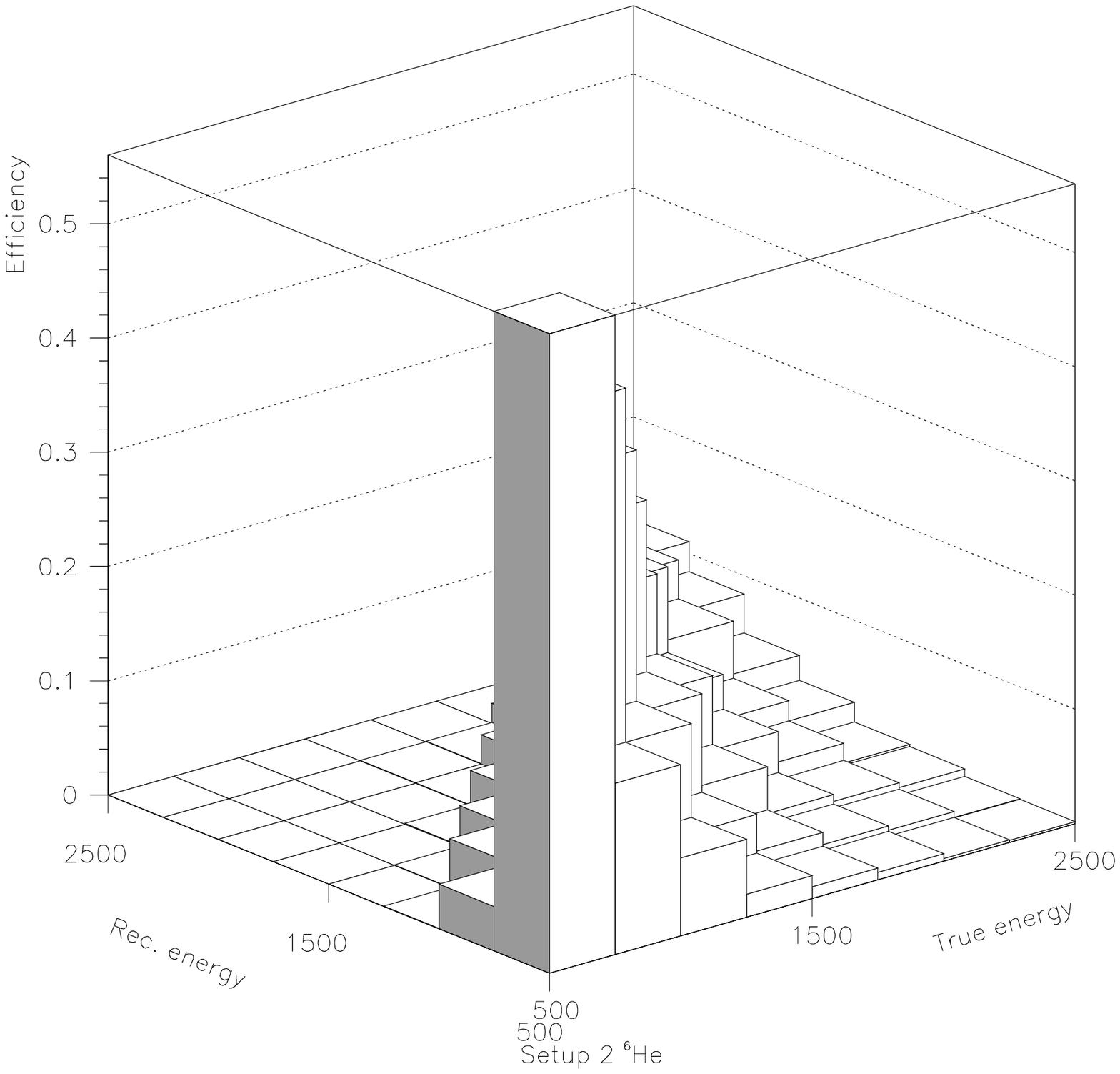,width=7cm,height=7cm} 
\epsfig{file=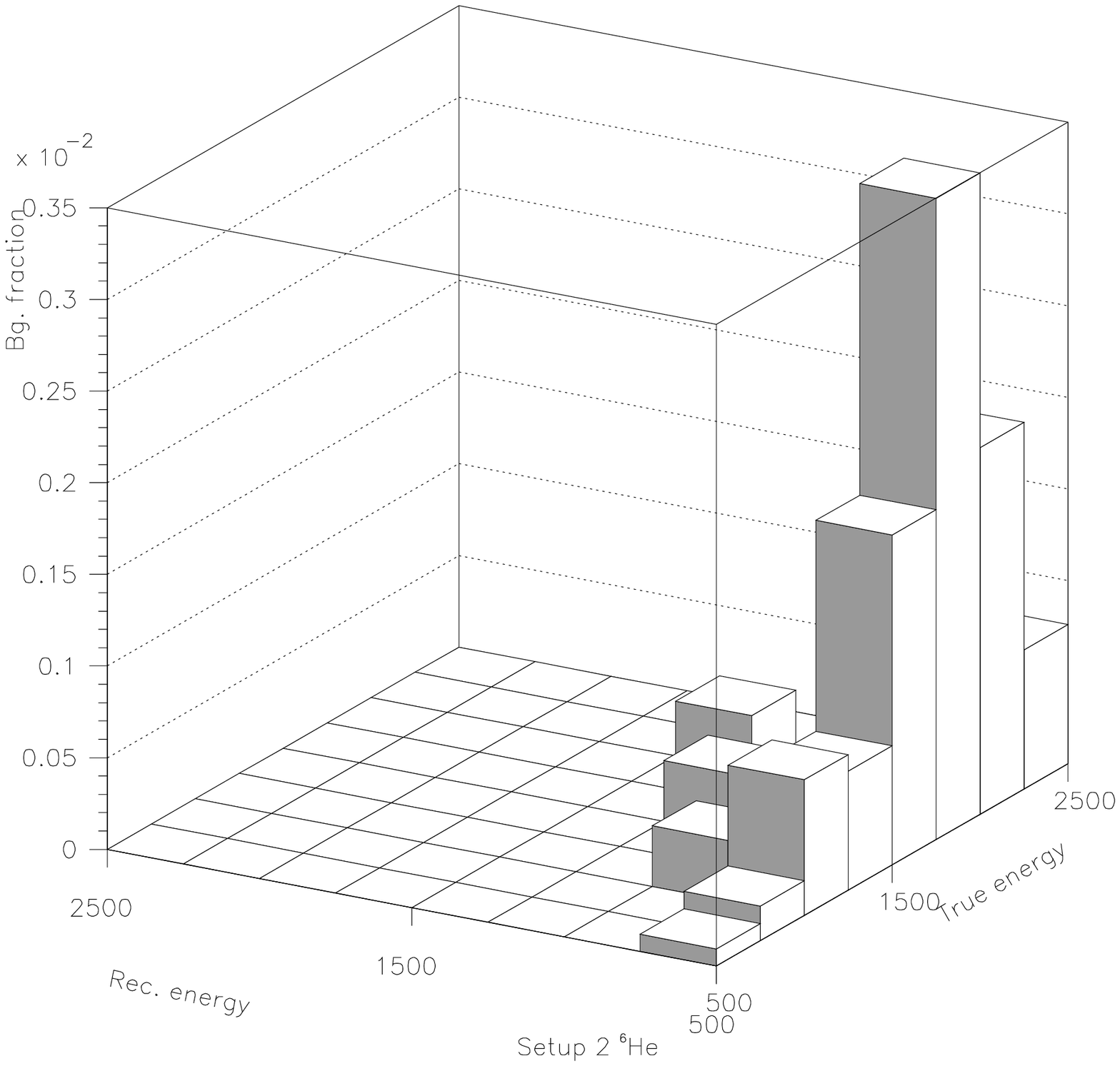,width=7cm,height=7cm} 

\caption[a]{Efficiencies as a function of the true and reconstructed 
neutrino energies for $^6$He in water in Setup II.}
\la{fig:effbgskhe6}

\end{figure}

One possibility to control still better the backgrounds 
would be to have a tunable $\gamma$. In this way
one could characterize the signal at a given energy reducing the background 
coming from higher energy events maximally. The optimization of such a 
design (how many years one should run at each $\gamma$) 
is an interesting one that will be considered elsewhere.

\begin{figure}[t]

\epsfig{file=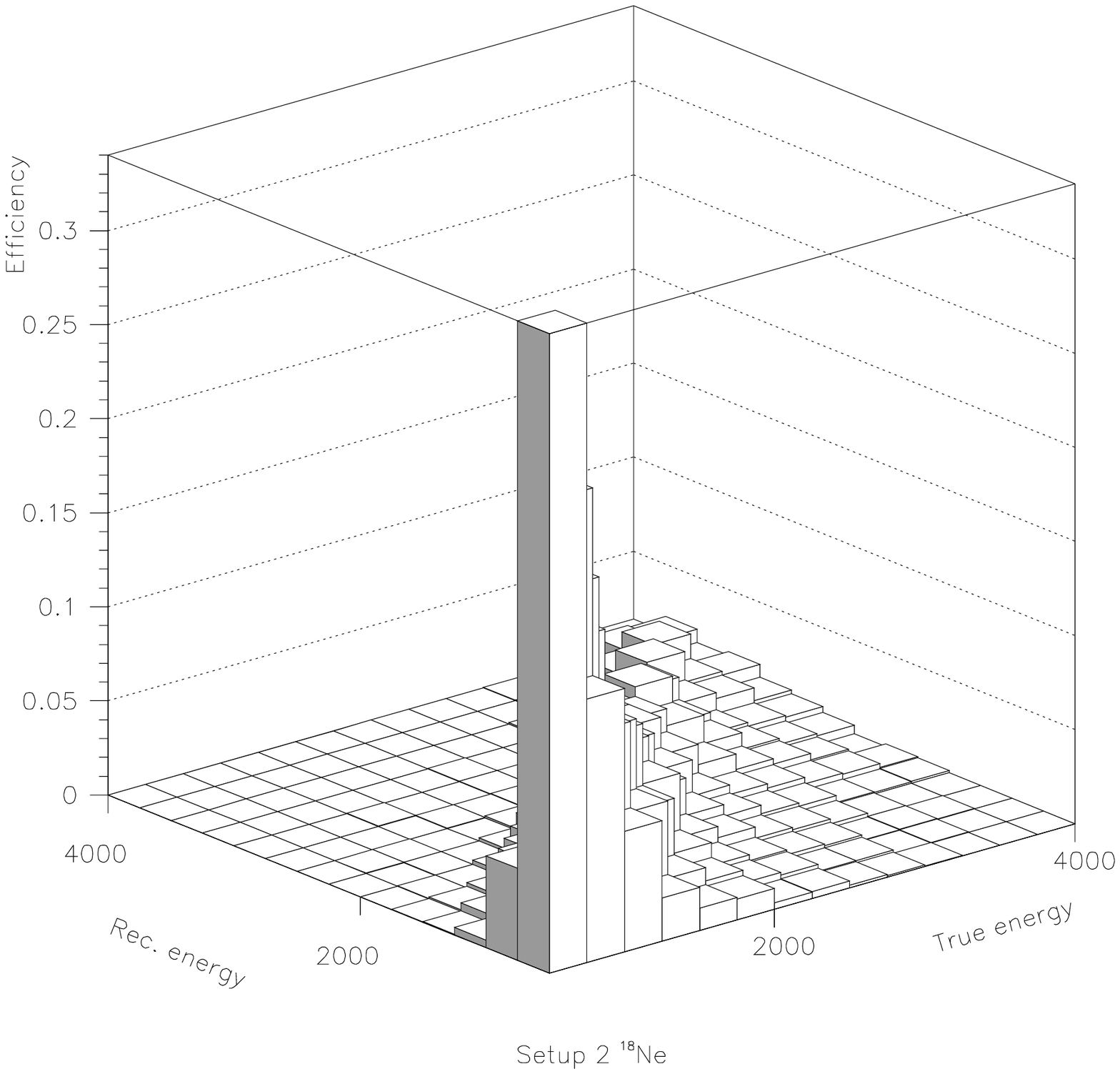,width=8cm,height=7cm} 
\epsfig{file=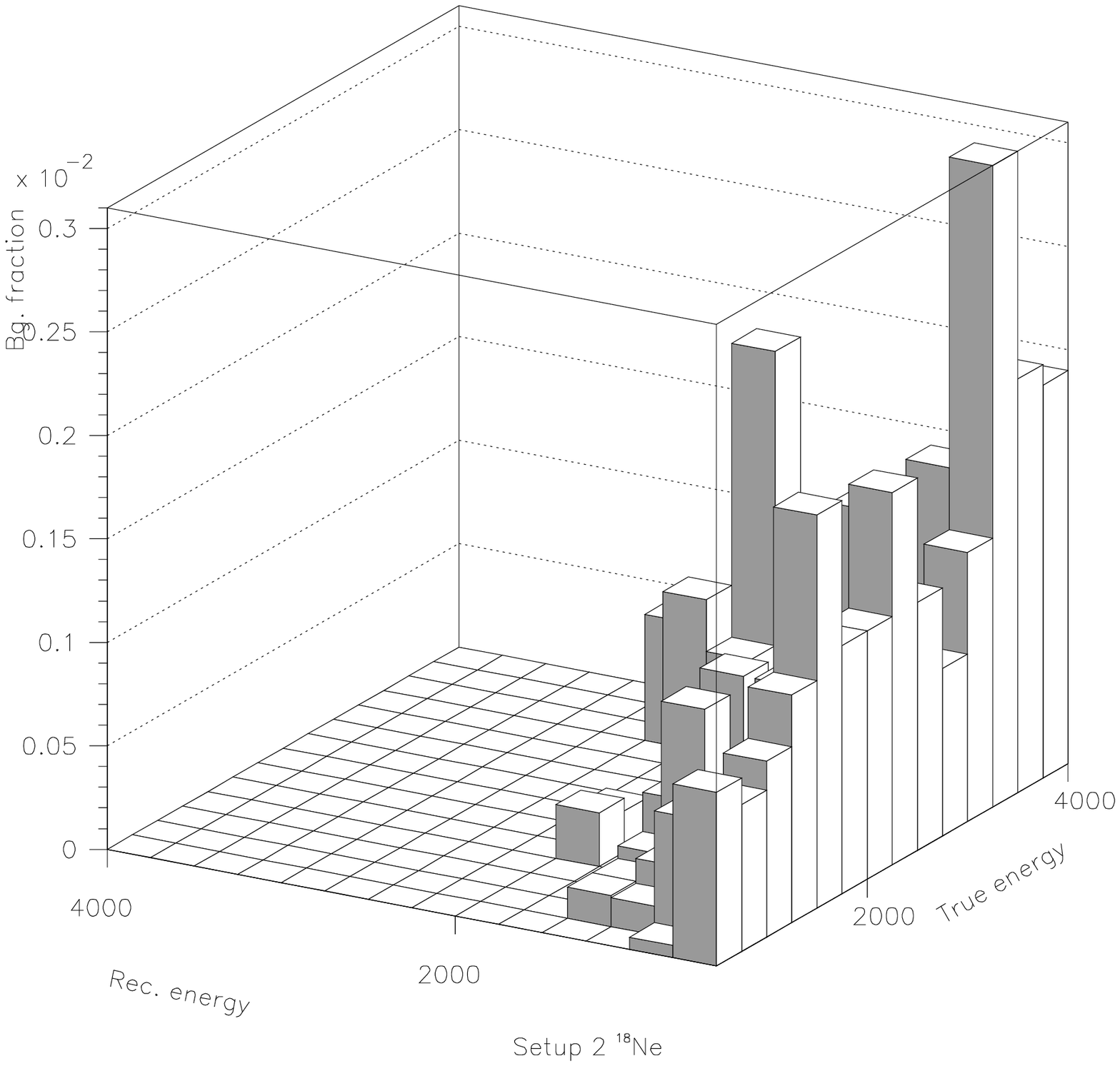,width=8cm,height=7cm} 

\caption{Efficiencies as a function of the true and 
reconstructed neutrino energies for $^{18}$Ne in water for the case
of Setup-II.}
\la{fig:effbgskne18}

\end{figure}

\subsection{Setup-III}
\label{tc}
The neutrino spectra for setup-III extends up to a few GeV, 
well in the CC regime, where
a tracking calorimeter (possibly a massive version of Minos \cite{minos}) could
offer better performance than water. The performance of such a device
(a 40 kiloton magnetized calorimeter) for the
case of the neutrino factory has been 
extensively studied \cite{cervera,golden}. However,
the neutrino energy for that setup was higher 
(mean energy of about 25 GeV to compare
with mean energy of about 5 GeV here) and the charge of the muon had to be
measured, which is not the case here. We have
not yet optimized the detector characteristics 
for the high energy option of the $\beta$-beam,
but we expect a similar performance, with efficiencies better than 
30 $\%$ and background fractions better than $10^{-4}$. 
This will be the subject of a forthcoming
study. In this paper, we will assume these numbers. 
 We also assume that, as it was the case in \cite{golden}, 
the neutrino energy can be reconstructed also for CC events. Energy 
bins of $1~$GeV will be considered and we discard events
with neutrino energies below $1$GeV.

%
\section{Determination of $\theta_{13}$ and $\delta$}

The simultaneous measurement of $\theta_{13}$ and $\delta$ 
is affected by correlations \cite{golden} and the so called
{\it intrinsic} degeneracy \cite{burguet1}, which 
results in either a proliferation of disconnected regions
of parameter space, where the oscillation probabilities are
very similar to be distinguished, or artificially large 
uncertainties in both parameters when these regions overlap. 

As has been discussed before, these degeneracies can be resolved
either combining different experiments with different $E_\nu/L$ 
or matter effects, or exploiting, whenever this is possible, 
the energy dependence of the signal within one experiment. 

One of the main advantages of going to higher energies and longer baselines
in the $\beta$-beam is precisely to have some significant energy 
resolution which allows to resolve these degeneracies. 

In Figure~ 
\ref{fig:exclu} we compare the reach concerning CP-violation on the 
plane $(\theta_{13},\delta)$, i.e.
the range of parameters where it is possible  
to distinguish with a $99\%$CL, $\delta$ from 0 or $180^\circ$ for the 
different setups and 10 years of running in each case. We
assume that both ions are bunched and accelerated simultaneously. We 
thus include the results from the measurement of both 
polarities.   
The remaining oscillation parameters are fixed close to their best fit values:
$\Delta m^2_{23}= 2.5 \times 10^{-3}$eV$^2$, $\theta_{23}=45^\circ$, 
$\Delta m^2_{12} = 7 \times 10^{-4}$eV$^2$, $\theta_{12}=35^\circ$. 

\begin{figure}[t]
\begin{center}

\epsfig{file=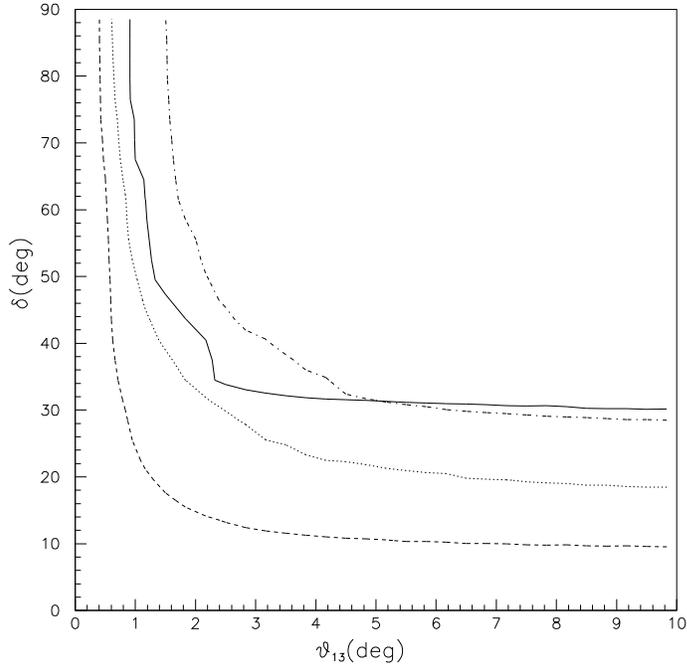,width=10cm,height=10cm} 

\caption[a]{Region where $\delta$ can be distinguished from $\delta=0$ or 
$\delta=180^\circ$ with a 99$\%$ CL for setup I (solid), setup II with 
the UNO-type detector of 400 kton described in section~\ref{uno} (dashed) and with the same detector with a factor 10 smaller mass (dashed-dotted) and setup III (dotted) with a 40 kton tracking calorimeter described in section~\ref{tc}. In all cases we consider 10 years of running time for both polarities (since they
are accelerated simultaneously).}
\la{fig:exclu}

\end{center}
\end{figure}
 
The solid line corresponds to the setup I, which has been studied before
\cite{mauro,blm}. The dashed lines correspond to setup II for the UNO detector 
described in the previous section and for a detector scaled down by a factor of 10. Surprisingly 
the small water Cerenkov in setup II performs similarly to the UNO detector in setup I, while the performance of the latter in setup II is spectacular and 
clearly comparable with the neutrino factory. One of the 
reasons for this improvement is precisely the resolution of correlations. 
This can be seen in Fig.~\ref{fig:fits1}, where we compare the result of the 
fits for setups I, II and III. While the 
intrinsic degeneracies are present for the low-energy option, they tend
to get resolved in the higher one, even with the smaller detector. 

Notice that the necesity to supress backgrounds due to charge
missidentification forces a very stringent cut in the 
momentum of the muon when searching for ``wrong sign'' muons at the
neutrino factory \cite{golden}. This cut translates in practice in
throwing away neutrino energies below ~ 5 GeV, thus loosing precious
spectral information. In that respect, the presence of two neutrino
species in the neutrino factory is a disadvantage, compared with the
$\beta-$beam, where one has only one neutrino species {\it and} 
the ability to identify low energy muons in water, 
separating them clearly from
backgrounds (as opposed to a tracking calorimeter, where a muon 
of momentum less than about 1 GeV cannot be easily separated from
the pion background).
 
Comparing setups II and III we realize that no improvement
is gained at the highest $\gamma$ as regards CP violation. The reason for
 this is probably the fact
that matter effects become dominant 
at this longer baselines (see Fig.~\ref{matter})
and hide 
the genuine CP violation. Although an optimization of the detector design
has not been performed in this case, it is clear that 
no gain will result with respect to setup II.

\begin{figure}[t]
\begin{center}

\epsfig{file=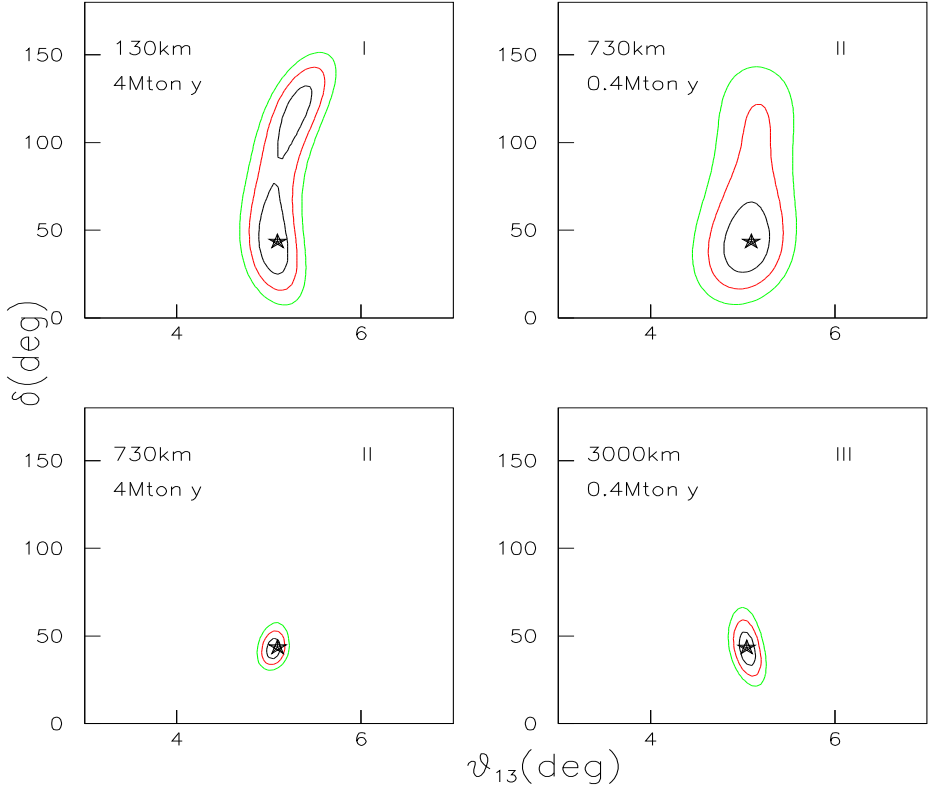,width=14cm,height=12cm} 

\caption[a]{1,2,3$\sigma$ contours on the plane $(\theta_{13},\delta)$ in the setups I, II for the $40$kton and $400$~kton detectors and for setup III in 10 years of running time. The input values of the parameters are indicated
by a star.}
\la{fig:fits1}

\end{center}
\end{figure}


Although other systematic errors, such as the knowledge of the flux or
the error on the backgrounds and efficiencies have not been included in this 
study, they are very unlikely to change the conclusion concerning the comparison of the three setups of the $\beta$-beam,  since they 
would affect them in a similar manner. However {\it all} systematic errors
should 
be included in a fair comparison of the $\beta$-beam and the neutrino factory, since they might be quite different in both machines.

\section{Determination of the sign$(\Delta m^2_{23})$}

The sign$(\Delta m^2_{23})$ is an essential missing piece of information 
to determine the structure of the neutrino mass matrix. The measurement
of this quantity cannot be done from the measurement of neutrino oscillations 
in vacuum, so matter effects need to be sizable. In setup I, matter
effects are too small to allow the determination of this unknown, however
this is no longer the case for the intermediate baseline setup. 

In Figure~\ref{fig:sign} we show the exclusion plot for the 
sign of $\Delta m^2_{23}$ on the plane $(\theta_{13},\delta)$ at $99\%$CL.
The measurement of the sign is possible in a very significant region of 
parameter space. In particular for the largest detector, it can be
measured for $\theta_{13} \geq 4^\circ$, simultaneously with 
$\theta_{13}$ and $\delta$ ! On the other hand, in setup III the sign can 
be measured in a larger region of parameter space. As discussed in the previous section this 
comes with the price that the sensitivity to CP violation gets spoiled.

\begin{figure}[t]
\begin{center}

\epsfig{file=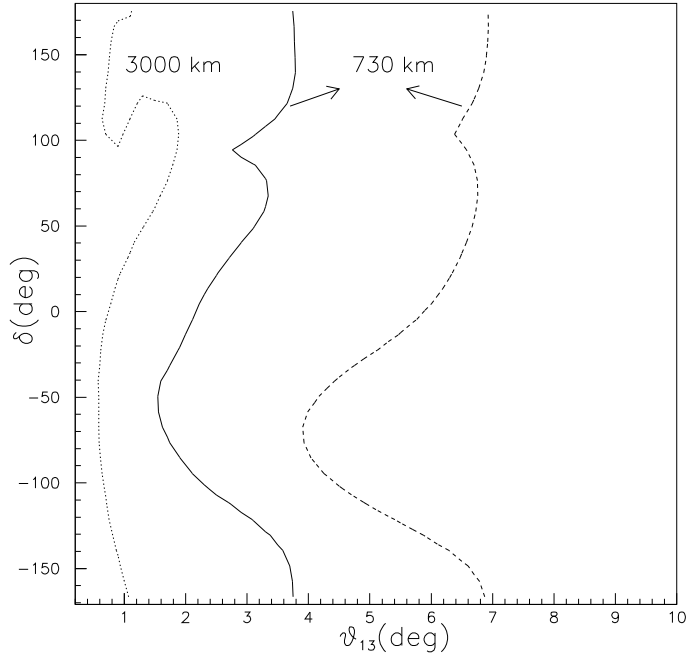,width=10cm,height=10cm} 

\caption[a]{Regions where the true sign$(\Delta m^2_{23})=+1$ can be measured 
at 99$\%$ CL (ie. no solution at this level of confidence
exists for the opposite sign). The lines correspond to setup II with the $400$kton water Cerenkov (solid), the $40$kton one (dashed) and to setup III (dotted) in 10 years of running time.} 
\la{fig:sign}

\end{center}
\end{figure}

%
\section{Conclusions and Outlook}

In this paper we have studied the physics potential of a higher $\gamma$ 
$\beta$-beam and compared it with that of the present design ($\gamma \sim 100$, $L =130$km). From a tecnical point of view, there does not seem 
to be any major difficulty in increasing the $\gamma$ of the $\beta$-beam.

From the physics point of view the 
increase in $\gamma$ is advantageous for three reasons: at the first 
atmospheric peak (which fixes $\gamma/L$)  the rates
increase linearly with $\gamma$, the increase in the baseline 
gives sizable matter effect, which allows to distinguish the sign of
$\Delta m^2_{23}$, and finally the energy dependence of the oscillation signal is easier to measure and gives precious information to resolve correlations
and degeneracies between oscillation parameters. 

From the experimental point of view, a very important advantage of the low-energy option was the possibility to use a very massive water Cerenkov a la UNO, remains true if the 
$\gamma$ is increased to $\gamma \sim 500$: the result of our realistic 
simulation shows that backgrounds can be reduced to a very small level
 with rather high efficiencies. At even higher energies 
water Cerenkov's are no longer optimal, but one could give up the increase
in statistics by decreasing the detector mass, in such a way that 
a different technology, such as a calorimeter, can be used.
 
We have thus considered two reference 
higher $\gamma$ options: $\gamma \sim 500$ and a 
water Cerenkov a la UNO at $730$km and $\gamma \sim 2000$ with a 10 times smaller calorimeter (which is still to be optimized) at 3000km. Our results show that the intermediate $\gamma$ option is spectacularly better than the low $\gamma$ option 
previously considered, both in terms of the reach in CP violation as in the 
possibility to measure the neutrino mass hierarchy. The highest $\gamma$ option
instead has an intermediate performance as regards the reach in CP violation. 
This is due to the fact that the baseline is so large that matter effects become dominant and hide to a large extent 
the genuine CP violation. On the other hand, the determination of the 
neutrino mass hierarchy is possible in a much enlarged region of parameter 
space. 

In this study we have compared the different $\gamma$ options alone. No combinations with possible super-beams or among themselves have been considered. Neither
the interesting possibility to get the silver transition $\nu_e \rightarrow \nu_\tau$ in the highest energy options has been studied. The combination of golden
and silver transitions has been shown to be extremely powerfull \cite{silver}
in resolving degeneracies so this is an interesting question that will be 
addressed elsewhere.

In summary, we have shown that the perspectives 
as regards the physics reach of the $\beta$-beam are
much more promising than previously thought, to the extent
that it can become competitive with the neutrino factory \footnote{A fair comparision between the $\beta$-beam and the neutrino factory should however
include a proper treatment of {\it all} systematic errors.}. In the light
of these results, we believe that the high-$\gamma$ option of a
$\beta$-beam design deserves careful consideration. Note in particular that
a $\beta$-beam, irrespective of the $\gamma$, does not require a proton
driver with so much power as that required by the SPL super-beam or 
the neutrino factory.  

Clearly a finer optimization of the $\gamma$(s) and baseline(s) 
to define an optimal roadmap for the $\beta$-beam complex
requires a more detailed study (and also the better knowledge of the 
atmospheric
and solar parameters that will be achieved in the next generation of 
neutrino experiments), but our results show that a $\gamma$ in the range of ${\cal O}(500)$ 
with a megaton water Cerenkov at a distance ${\cal O}(1000$km) will be hard
to beat.

%
\section*{Acknowledgement}

We warmly thank A.~Donini and S.~Rigolin for usefull discussions and 
also M.~Lindroos for his ideas on the feasibility 
of a higher energy $\beta$-beam and his encouragement. 
This work has been partially supported 
by FPA2002-00612 and FPA2001-1910/C03-02 of the CICYT and GV00-054-1.

\newpage


\appendix
\renewcommand{\thesection}{Appendix~\Alph{section}}
\renewcommand{\thesubsection}{\Alph{section}.\arabic{subsection}}
\renewcommand{\theequation}{\Alph{section}.\arabic{equation}}

\newpage


\end{document}